# Cationic vacancy induced room-temperature ferromagnetism in transparent conducting anatase Ti$_{1-x}$Ta$_x$O$_2$ (x~0.05) thin films

A. Rusydi[abc*], S. Dhar[ad*], A. Roy Barman[ab*], Ariando[ab], D.-C. Qi[ab], M. Motapothula[ab], J.B. Yi[e], I. Santoso[ab] Y. P. Feng[ab], K. Yang[bf], Y. Dai[f], N. L. Yakovlev[g], J. Ding[ae], A.T.S.Wee[ab], G. Neuber[h], M. B. H. Breese[abc], M. Ruebhausen[ah], H. Hilgenkamp[ai], and T. Venkatesan[abd]

[a]*NUSNNI-NanoCore, National University of Singapore, Singapore 117411, Singapore.*

[b]*Department of Physics, National University of Singapore, Singapore 117542, Singapore.*

[c]*Singapore Synchrotron Light Source, National University of Singapore, Singapore 117603, Singapore.*

[d]*Department of Electrical and Computer Engineering, National University of Singapore, Singapore 117576, Singapore.*

[e]*Department of Materials Science & Engineering, National University of Singapore, Singapore 117576, Singapore.*

[f]*School of Physics, State Key Laboratory of Crystal Materials, Shandong University, Jinan 250100, P. R. China.*

[g]*Institute of Materials Research and Engineering, A*STAR, Singapore 117602, Singapore.*

[h]*Institut für Angewandte Physik, Universität Hamburg, D-20355 Hamburg, Germany. Center for Free Electron Laser Science (CFEL), D-22607 Hamburg, Germany.*

[i]*Faculty of Science and Technology and MESA+ Institute for Nanotechnology, University of Twente, P.O. Box 217, 7500 AE Enschede, Netherlands.*

Corresponding authors. Email: T.V. (venky@nus.edu.sg) or A.R. (phyandri@nus.edu.sg) or S.D. (eleds@nus.edu.sg)

[*]These authors contributed equally to this work.

We report room-temperature ferromagnetism in highly conducting transparent anatase $Ti_{1-x}Ta_xO_2$ (x~0.05) thin films grown by pulsed laser deposition on $LaAlO_3$ substrates. Rutherford backscattering spectrometry (RBS), x-ray diffraction (XRD), proton induced x-ray emission (PIXE), x-ray absorption spectroscopy (XAS) and time-of-flight secondary ion mass spectrometry (TOF-SIMS) indicated negligible magnetic contaminants in the films. The presence of ferromagnetism with concomitant large carrier densities was determined by a combination of superconducting quantum interference device (SQUID) magnetometry, electrical transport measurements, soft x-ray magnetic circular dichroism (SXMCD), XAS, and optical magnetic circular dichroism (OMCD) and was supported by first-principle calculations. SXMCD and XAS measurements revealed a 90% contribution to ferromagnetism from the Ti ions and a 10% contribution from the O ions. RBS/channelling measurements show complete Ta substitution in the Ti sites though carrier activation was only 50% at 5% Ta concentration implying compensation by cationic defects. The role of Ti vacancy and $Ti^{3+}$ was studied via XAS and x-ray photoemission spectroscopy (XPS) respectively. It was found that in films with strong ferromagnetism, the Ti vacancy signal was strong while $Ti^{3+}$ signal was absent. We propose (in the absence of any obvious exchange mechanisms) that the localised magnetic moments, Ti vacancy sites, are ferromagnetically ordered by itinerant carriers. Cationic-defect-induced magnetism is an alternative route to ferromagnetism in wide-band-gap semiconducting oxides without any magnetic elements.

# 1.    Introduction

Because of many potential applications in spintronics, magneto-optic, and opto-electronic devices, extensive efforts are being made to introduce room-temperature (RT) ferromagnetism (FM) into various wide-band-gap semiconducting oxides and III-V and II-VI based semiconducting systems [1-15]. Considerable success has been made in the production of dilute magnetic semiconductors (DMS) by introducing a magnetic impurity into a non-magnetic host material. The most reproducible carrier-induced FM in semiconductors is based on Mn-doping of GaAs, but its Curie temperature $T_c$ is limited to about 173 K [3.7]. Among the oxides, the most notable compounds exhibiting DMS behaviour above RT are based on $TiO_2$ and ZnO [8-15]. Despite these developments, magnetic-impurity-based DMS oxides are a subject of intense debate due to the many inconsistent results related to secondary phase formation, unknown impurities or substitutionality, solubility, clustering and/or segregation of the magnetic impurities in the host oxide matrix [16-18]. More recently, there have been reports of FM in ZnO and other oxides (attributed to anionic vacancy) by doping with non-magnetic ions, but besides evidence from SQUID data [13-15], a necessary but not sufficient criterion [17-18], there has been no reliable element-specific evidence for FM in these cases with the exception of recent publications [19-20]. To establish true FM in the DMS oxides and to differentiate from impurity artefacts, various magnetic measurements, ranging from extremely sensitive SQUID and OMCD measurements, which directly probe the spin-polarised bands, to element-specific SXMCD measurements, are required.

The idea of cationic-vacancy-induced FM (with concomitant half-metallicity) in wide band-gap semiconducting oxides was proposed on theoretical grounds by Elfimov *et al*. [21] and later Osorio-Guillén *et al*. [22] pointed out that either hole or electron doping facilitates the formation of cationic vacancies. It was predicted that wide band gap CaO could become a half-metallic ferromagnet with about 3-5% cationic (Ca) vacancies. However, the realisation of high-quality films of CaO has not yet been achieved due to the hygroscopic nature of CaO, which leads to rather unstable films. The first experimental observation of a local magnetic moment arising from cationic vacancy was reported by Zhang *et al*. [23], who observed signatures of Kondo effect below 100 K in 5% Nb doped anatase $TiO_2$ thin films grown under $10^{-4}$ Torr oxygen partial pressure by pulsed laser deposition (PLD). Using XAS and XPS, supported by first-principle calculations, they had shown that the appearance of Kondo scattering was due to the presence of localised magnetic moments associated with cationic (Ti) vacancies produced due to Nb incorporation. However, no FM was observed presumably due to the low concentration of Ti vacancies and free carriers [8]. On the other hand from the structural consideration, Osorio-Guillén *et al.* [22] suggested the difficulty of creating films having cationic vacancies larger than 3%. In this paper we report the production of Ti vacancies ($V_{Ti}$ ) and room temperature FM over a narrow processing range in (001) anatase $Ti_{1-x}Ta_xO_2$ (x~0.05) thin films grown on (001) $LaAlO_3$ substrates. We show that $Ta^{5+}$ gives rise to free charge carriers and cationic defects i.e. $V_{Ti}$ and $Ti^{3+}$ are created as compensating centers depending on the processing condition. From XMCD, relative contributions to FM from cationic versus anionic defects are in the ratio of 9 to 1. The appearance of FM is proposed to be due to the formation of Ti vacancies (with delocalized $V_{Ti}$ magnetic orbitals) coupled by the itinerant electron-carrier-mediated (RKKY) exchange interaction.

## 2. Experimental Method

Very high-purity (99.999%) $Ta_2O_5$ and $TiO_2$ powders were ground for several hours before sintering in a furnace at 1000°C in air for 20 hours. Next, target pellets were made and sintered at 1100°C in air for 24 hours. Anatase $Ti_{1-x}Ta_xO_2$ epitaxial thin films (with $x = 0.00$ and 0.1) of thicknesses between 50 and 500 nm were deposited on high-quality (001) $LaAlO_3$ substrates by PLD using a 248-nm Lambda Physik (Germany) excimer laser with an energy density of 1.8 J/cm$^2$ and a repetition rate of 2-10 Hz. Depositions were performed for 0.5-1 h in a stable oxygen partial pressure of $1\times10^5$ Torr while the substrate temperature was maintained at various temperatures in the range of 500-800 °C. The chemical and structural properties of the samples were studied by XPS, electrical transport measurements, SQUID, XAS, SXMCD, OMCD, RBS-Channeling, XRD and TOF-SIMS.

## 3.  Results and discussion

In this study, we found that the preferred temperature of growth for a highly crystalline pure $TiO_2$ films was about 750 $^o$C and observed a catalytic effect of Ta on $TiO_2$ which reduced the crystallization temperature with increasing Ta concentration. While FM was demonstrated on many samples, we present here four types of films to to illustrate the role of $V_{Ti}$ and $Ti^{3+}$ on the origin of FM in this system: pure $TiO_2$ films grown at 600 and 750 $^o$C, and $Ti_{1-x}Ta_xO_2$ (x~0.05) films grown at 600 and 750 $^o$C.

## 3.1 Structural, chemical, electrical and optical properties

The formation of c-axis-oriented anatase $TiO_2$ was confirmed in all $Ti_{1-x}Ta_xO_2$ films by XRD analysis, and no other phases were observed within its detection limit. Using ultra violet-visible (UV-VIS) spectroscopy, the optical band-gaps of the pure and doped samples were found to be 3.32 and 3.4 eV, respectively. The actual Ta concentrations in the films were found to be about 5.5±0.3% by RBS (Figure 1) and the ion channelling spectra show that most (99%) of the Ta was substitutional and likely in the Ti sites.

The contents of magnetic trace elements were measured using TOF-SIMS with 25-kV Bi analysis ions and 3 kV Ar sputtering ions. Because the main aim of this measurement was to determine the content of magnetic impurity elements, a reference target containing 1 at. % each of Cr, Mn, Fe, Co and Ni and 95% of Ti, as measured by the weight of the respective oxides, was prepared. A film from this calibration target was made under the same conditions used to produce the Ta substituted $TiO_2$ ferromagnetic films. The intensities of the various elemental peaks in the SIMS mass spectra were integrated and normalised to that of $^{46}Ti$. The relative sensitivity factors of these elements were determined from their intensities in the reference sample and then the contents of all elements were estimated (Figure 2). Unintended magnetic impurities (Fe, Ni, Mn, Co, Cr) were found to be substantially below 0.01% (about three orders of magnitude less than the Ta concentration and roughly five orders of magnitude lower than the Ti concentration) in all samples by RBS (inset of Figure 1), TOF-SIMS and wide-scan of XAS (not shown here). Therefore, the polarising effect (a possibility) of the magnetic impurities on Ti ions can be clearly ruled out.

XPS data in Figure 3 show peaks at 459.1 eV (Ti $2p_{3/2}$) and 464.8 eV (Ti $2p_{3/2}$) which originate from $Ti^{4+}$, while peaks at 457.3 and 462.4 eV are from $Ti^{3+}$ confirming that Ti is dominantly in the 4+ state for both pure $TiO_2$ and $Ti_{1-x}Ta_xO_2$ (x~0.05) samples. Our detailed analysis suggested that $Ti^{3+}$/$Ti^{4+}$ ratios for pure $TiO_2$ (Figure 3(a)) and $Ti_{1-x}Ta_xO_2$ (x~0.05) (Figure 3(b)) films grown at 600 $^o$C were ~0.65% and ~0.59%, respectively, while it was ~8.45% for the $Ti_{1-x}Ta_xO_2$ (x~0.05) films grown at 750 $^o$C (Figure 3(c)) [24]. The Ta was found to be in the 5+ state in all the films [25]. It may be noted that $Ti^{4+}$ state is non-magnetic while $Ti^{3+}$ state is magnetic in nature.

The electron carrier concentration of the *n*-type pure anatase $TiO_2$ sample was of the order of $10^{17}$ to $10^{19}$ /cm$^3$. After 5 at.% Ta incorporation, the *n*-type $Ti_{1-x}Ta_xO_2$ (x~0.05) films became highly conductive. The room temperature electron carrier density and Hall mobility of $Ti_{1-x}Ta_xO_2$ (x~0.05) films grown at 600 °C were found to be about $7.66 \times 10^{20}$ /cm$^3$ and 13.7 cm$^2$/Vs, respectively, while that for the 750 °C films they were $4.90 \times 10^{20}$ /cm$^3$ and 11.9 cm$^2$/Vs, respectively. The full carrier activation was incomplete for all the Ta concentrations implying carrier compensation which peaked at about 5-6% Ta concentration [26].

## 3.2 Magnetic properties

To study the ferromagnetic properties, SQUID measurements were performed on the pure and Ti$_{1-x}$Ta$_x$O$_2$ (x~0.05) thin films in a number of samples grown under different Ta concentrations, temperatures and oxygen partial pressures. In a study involving more than 60 samples, the magnetisation tended to peak at a Ta concentration of about 5-6%, deposition temperature of about 600 °C (repeated for about 20 samples) and an oxygen pressure of about $10^{-5}$ Torr.

A large number of samples prepared at optimum condition showed FM with saturation magnetization ranging from 1 to 4 emu/g with coercive fields ranging from 70 to 90 Oe. Efforts are being made to narrow down the scatter in the saturation magnetization value. In order to obtain a better signal to noise ratio in other measurements, we used samples that showed saturation magnetization values close to 4 emu/g. A magnetization curve for a Ti$_{1-x}$Ta$_x$O$_2$ (x~0.05) film grown at 600 °C along with the one obtained from a pure TiO$_2$ film (non-magnetic) is shown in Figure 4. In Ti$_{1-x}$Ta$_x$O$_2$ (x~0.05) film grown at 600 °C, the magnetization of 4 emu/g implies 1.1 $\mu_B$/Ta using 5.5% Ta ions (obtained from the RBS analysis), however, later analysis will reveal that Ta is not directly responsible for the FM but rather Ti vacancies induced by it. In the same figure the magnetization for a higher crystallinity Ti$_{1-x}$Ta$_x$O$_2$ (x~0.05) film grown at 750 °C is also shown, with magnetization values less by a factor of ~20 suggesting the possible role of defects in the FM. The LaAlO$_3$ substrate treated under similar processing conditions (oxygen pressure and temperature but no film deposition) showed only diamagnetic behaviour. The Curie temperature T$_c$ of the sample was well above 100 °C, above which reliable measurements are hard due to gradual modification of the material.

Although the SQUID data clearly show the presence of RT FM in the Ti$_{1-x}$Ta$_x$O$_2$ (x~0.05) films (prepared at 600C) , they do not provide microscopic insight into the origin of the ferromagnetic ordering. Thus, SXMCD measurements were performed in total yield mode as a function of photon energy using elliptically polarised light with a degree of circular polarisation $p$ = 90 and an energy resolution of 0.25 eV at the SINS beam line of the Singapore Synchrotron Light Source (SSLS) [27] because this technique is not only element-specific but also capable of estimating both the spin and orbital magnetic moments and their anisotropies [28-30]. To measure the in-plane magnetic moments, the light was incident at a grazing angle of 20° from the sample surface, with its propagation direction along sample in-plane magnetisation direction. An external saturation magnetic field of up to 2000 Oe was applied to magnetise the sample along the in-plane direction. The SXMCD is the difference between two XAS spectra in the soft energy range taken with different light polarisations or different magnetic field directions. To avoid any ambiguity, both methods were applied. In the first method, the XAS data were taken with two opposite circular polarisations relative to a fixed sample magnetisation direction (this was achieved by turning on a 0.2-T field for a short duration). The absorption coefficients, $\mu^+$ and $\mu^-$, which are directly proportional to XAS, have a photon helicity (spin) direction ($\mu^+$) and anti-parallel ($\mu^-$) to the sample magnetisation and magnetic field (M, H) direction, respectively [27-28]. Therefore, the SXMCD is equal to ($\mu^- - \mu^+$). In the second method, the XAS data were taken with two opposite sample magnetisation directions while fixing one of the directions of the circular polarised light. It is important to note that these two methods yield very similar results, indicating that the SXMCD results directly reflect the intrinsic properties of films. The detection mode for both the SXMCD and XAS measurements was total electron yield (probing <20 nm) and fluorescence yield (probing <200 nm). Because the signal-to-noise ratio of the electron yield was superior, we only show here the total electron yield data.

The SXMCD data of $Ti_{1-x}Ta_xO_2$ (x~0.05) films grown at 600 $^o$C and 750 $^o$C taken without any external magnetic field are shown Figure 5. In order to compare the SXMCD results at various edges, the SXMCD data were normalized to an absolute scale by fitting to the Henke tables [31] far below and above the edges, and then the absorption coefficients, $\mu^+$ (parallel) and $\mu^-$ (antiparallel) were calculated. The advantage of this procedure is that it can be applied to most of resonant edges in the soft X-ray range and thus allows us to comapre the absorption coefficients at various edges [32-35], i.e. the Ti $L_{2,3}$ edges ($2p \rightarrow 3d$ transitions) and the O $K$ edge (O $1s \rightarrow 2p$ transition) for this study. The $\mu^+$ and $\mu^-$ at the Ti $L_{2,3}$ edge consist of two sets of peaks separated by 5-6 eV due to core hole spin-orbit coupling [36-37] of Ti $2p_j$ with $j$=1/2 or 3/2. Moreover, due to ligand-field splitting, the $3d$ bands can be identified as $t_{2g}$- and $e_g$- symmetry bands. As a result, the Ti $2p \rightarrow 3d$ transitions consist of 4 dominant structures and all the relevant transitions are shown in the attached Table 1. It may be noted that the Ta $L_{3,2}$ edges are expected to occur at around 9881 and 11136 eV which are out of the soft X-ray range [31]

The $\mu^+$ and $\mu^-$ at the Ti $L_{2,3}$ and O $K$ edges of the Ti$_{1-x}$Ta$_x$O$_2$ (x~0.05) sample are shown in Figure 5(a) and (b), respectively. The SXMCD spectra in Figure 5(c) and 4(d) correspond to the remnant magnetisation and are the most direct evidence for the intrinsic FM [17]. SXMCD signals are surprisingly robust despite the fact that there was no applied magnetic field during the measurement. The observation of remnant SXMCD signals rules out the possibility of super-paramagnetism [29-31]. In contrast, the pure TiO$_2$ film did not show any SXMCD signal at both resonant Ti $L_{2,3}$ and O $K$ edges. Further, higher crystallinity samples grown at 750 °C showed SXMCD (Figure 5(c)) signal ~20 times smaller consistent with the SQUID measurement.

*Table 1:* SXMCD peaks and corresponding transitions at the Ti $L$ and O $K$ edges

| Peak Position of SXMCD Signal | Elemental Edge | Transition |
|---|---|---|
| 458.2 eV | Ti | $2p_{3/2} \rightarrow 3d(t_{2g})$ |
| 460 eV | Ti | $2p_{3/2} \rightarrow 3d(e_g)$ |
| 463.5 eV | Ti | $2p_{1/2} \rightarrow 3d(t_{2g})$ |
| 465.5 eV | Ti | $2p_{1/2} \rightarrow 3d(e_g)$ |
| 530.6 eV | O | $1s \rightarrow$ Ti $3d(t_{2g})$ |
| 533.2 eV | O | $1s \rightarrow$ Ti $3d(e_g)$ |

The fact that both the Ti $L$ and O $K$ edge showed SXMCD signals dominantly at the $t_{2g}$-state indicates a strong $p$-$d$ hybridisation and suggests that the $t_{2g}$-derived state plays a dominant role in the observed FM in this system. By applying the X-ray MCD sum rule [28,29], we obtained that the contribution to the orbital magnetic moment of Ti was nine times stronger than that of O and the spin alignment at Ti and O was parallel. These data confirm unambiguously that the origin of the FM is related to Ti sites.

We further investigated the nature of this magnetism via OMCD, which is a photon-in and photon-out measurement, with an extended Sentech SE850 ellipsometer at the University of Hamburg, Germany. It covered an energy range from 0.5 to 5.5 eV and was equipped with an ultra high vacuum cryostat. For SGME, additional mounted Helmholtz coils enable the application of a fast-switching external magnetic field of 4500 Oe in TMOKE geometry [38]. As the penetration depth was above 200 nm in this energy range, the whole film was scanned optically.

Figure 6(a) displays the differential intensity change δI/I as a function of photon energy from 2 to 5 eV at various magnetic fields for the $Ti_{1-x}Ta_xO_2$ (x~0.05) film grown at 600 °C. The position of the measured optical band gap of 3.42 eV is shown by a vertical dotted line. It is seen from the figure that the transitions around 3.5 and 4.5 eV are strongly influenced by the applied magnetic field. The change in sign of ΔI/I at 4.5 eV is due to the optical transitions from non-spin-polarised occupied states to two possible unoccupied states: majority-spin states at $E_F$ and the minority-spin states roughly 1 eV above $E_F$. The spin-splitting energy between the up-spin (majority spin) and the down-spin (minority spin) and spin polarisation of carriers at the optical band gap was about 1 eV which is similar to those found in colossal magneto-resistive manganites [39] suggesting a strong electron localisation effect (in the energy or k space) in this present system. Assuming that the bands close to Fermi energy ($E_F$) have a low density of states, the width of the transition at 3.5 eV of about 0.7 eV corresponds to the width of the spin majority band. This strongly suggests that the occupied states correspond to O 2$p$ states, whereas the majority and minority spin unoccupied states correspond to Ti 3$d$ states. We next plot (Figure 6(b)) the integrated absolute value of the OMCD signal between 2.2-4.1 eV and between 4.2-4.75 eV as a function of the applied magnetic field going from 0 to 4500 Oe and back. The hysteresis loop obtained from the SQUID measurement overlaps quite well (inset of Figure 6(b)) with the OMCD data. It is worth noting that the OMCD signal is significant and indicates that we have an intrinsic ferromagnetism as it is also connected to the optical band gap i.e. the electronic density of states of the intrinsic band structure. These further support that the FM seen in all the magnetization measurements came from the same source.

## 3.4 Theoretical calculation

To get further insight, first-principle calculations within a spin-polarised generalised-gradient approximation plus the on-site $U$ parameter (GGA+$U$) were performed. A 48-atom super-cell, modelled by 2×2×1 repetition of the 12-atom conventional unit cell of anatase $TiO_2$, which is proportional to 6.25% Ta doping, was employed to study the electronic structure of $Ti_{1-x}Ta_xO_2$ films. For computational difficulties, the 6.25%-Ta dopant was used instead of actual 5.5% Ta, but this does not change the main conclusions. The effective on-site $U$ parameter ($U_{eff}=U-J$) of 5.8 eV and a scissor operator were used to make the calculated band gap comparable with the experimental value. The calculated electronic structures are shown in Figure 7(a)-(d). Interestingly, the magnetic moment mainly resides at Ti 3d($t_{2g}$) bands which are hybridized with O $2p$ bands which is consistent with SXMCD results. Our calculations also indicated that an isolated $V_{Ti}$ produced a high-spin-polarization electronic state, which was mainly contributed by the O $2p$ orbital of the first-nearest O atoms around the $V_{Ti}$. Furthermore, the $V_{Ti}$ - $V_{Ti}$ interaction resulted in a stable ferromagnetic ground state. It is important to mention that the spin-polarized density induced by the $V_{Ti}$ extends very long to the third- and even to the fifth-nearest O atoms. As a result, the spin-polarized $V_{Ti}$ orbitals were delocalized. When the magnetic orbital of two $V_{Ti}$ overlap through the common spin-polarized, third- and fifth-nearest O atom, the overlapping spin density is nonzero, thereby leading to a long-range ferromagnetic alignment between the magnetic orbitals of two $V_{Ti}$ [40]. In this process, free electron carriers are expected to facilitate the spin exchange coupling interaction between them. Our calculations also indicate that the $Ti^{3+}$ - $Ti^{3+}$ interaction favours antiferromagnetism. This also fully supports our experimental observations in which sample with high $Ti^{3+}$, i.e. Ta-$TiO_2$ grown at 750 $^o$C Ta, has weak FM. Furthermore, the splitting energy between up-spin (majority spin) and down-spin (minority spin) Ti $3d$ states is about 1.1 eV. This value is also consistent with the OMCD measurements of $Ti_{1-x}Ta_xO_2$ (x~0.05) film grown at 600 $^o$C shown in Figure 6(a) with a measured spin-splitting energy of about 1 eV. The calculated OMCD based on DOS is compared with the experimental results for the $Ti_{1-x}Ta_xO_2$ (x~0.05) film grown at 600 $^o$C in Figure 8. Due to the quantum mechanical selection rules for optical absorption, the spin-polarised majority and minority states are directly connected with the appearance of a magneto-optical response. In the energy range of 2-5 eV, the optical transitions are dominated by charge transfer excitations between O $2p$ and Ti $3d$. The calculations considered the intersite transition from spin-up (spin-down)-occupied O $2p$ states to the spin-down (spin-up)-unoccupied Ti $3d$ states. Interestingly, the calculated OMCD spectrum tracks very well with the experimental data including the peak position and the width. This further supports that the FM is truly from intrinsic properties of the system.

## 3.5 Origin of Ferromagnetism

The evidence for FM from a variety of techniques is reassuring and what is left is to figure out the nature of cationic defect responsible for this. The Ti vacancy ($V_{Ti}$) and $Ti^{3+}$ are the likely candidates and we need spectroscopic evidence to identify their role. The XPS data in Figure 3 shows that the weakly ferromagnetic $Ti_{1-x}Ta_xO_2$ (x~0.05) film grown at 750 °C (best crystalline) exhibits 14 times higher $Ti^{3+}$ signal than the ferromagnetic $Ti_{1-x}Ta_xO_2$ (x~0.05) film grown at 600 °C. These data rule out the role of $Ti^{3+}$ in the FM seen.

The XAS spectra taken at the Ti $L_{2,3}$ edges from the pure and $Ti_{1-x}Ta_xO_2$ (x~0.05) films (both grown at 600 and 750 °C) are shown after background correction [32-36] in Figure 9(a) and (b) respectively. The XAS signal of $Ti_{1-x}Ta_xO_2$ (x~0.05) films grown at 600 °C increases dramatically in the $t_{2g}$ bands as compared to pure $TiO_2$ film grown at the same temperature . The increasing spectral weight is the direct evidence [23] for the formation of $V_{Ti}$ because a $V_{Ti}$ creates four holes in the O $2p$ band, which is strongly hybridised with the Ti $3d$ band. The creation of these holes increases the number of unoccupied states near the Fermi level, e.g., in the $t_{2g}$ bands and therefore increases the XAS signal (which is also consistent with our theoretical calculations shown above). From the ratio of the $t_{2g}$ to $e_g$ bands for the pure, $Ti_{1-x}Ta_xO_2$ (x~0.05) film grown at 600 °C (Figure 9(a)) we were able to estimate an upper limit of 3% for the $V_{Ti}$ with the actual number likely to be lower by a factor of two or more as it did not account for all the defects that could increase the number of unoccupied states near the Fermi level. This number is a factor of 5 larger than the one would arrive for the vacancy concentration based on charge compensation. Altogether, one can say for sure that the actual vacancy concentration is somewhere in between 0.6-3%. Much more detailed study would be needed to further narrow down these numbers. By the same analogy, when we compare the pure $TiO_2$ and $Ti_{1-x}Ta_xO_2$ (x~0.05) films grown at 750 °C (Figure 9(b)) with, the $t_{2g}$ peak of the later decreases in height with respect to the former (together with the support of XPS data) suggesting no or very reduced $V_{Ti}$.

In retrospect, it is important to note that the SXMCD signal in the present case is different from the one observed in $LaMnO_3/SrTiO_3$ originating from the $Ti^{3+}$ states present at its interface as recently reported by Garcia-Barriocanal $et$ $al$ [41]. This also suggests that the strong SXMCD signal in $Ti_{1-x}Ta_xO_2$ (x~0.05) films grown at 600 °C does not arise from the $Ti^{3+}$ defect and is most likely from the $V_{Ti}$. All these facts also support the fact that the role of $Ti^{3+}$, if any, is secondary.

Now that $V_{Ti}$ is established as the magnetic entity, we will try to develop a microscopic understanding of the FM. The four unpaired electrons in a Ti vacancy site can align in three possible ways which will yield 4, 2 and 0 $\mu_B$. Statistically we can assume a value of 2 $\mu_B$ per vacancy which would mean that to get the magnetization value seen, an amount of ~2.5% vacancy would be needed. In addition, to compensate 50% of the free electrons from the Ta, about 0.6% vacancies would be needed. So a total of about 3% $V_{Ti}$ is adequate to explain the saturation magnetization as well as the electron compensation seen which is consistent with the predictions [22]. The average distance between two Ti vacancies is about 3-4 unit cells. Unless the orbital magnetization of the Ti vacancy is extended over at least two unit cells, the direct exchange probability is very low. The fact that the FM (but instead see Kondo scattering [23]) not seen in samples prepared at higher oxygen pressures where the Ti vacancy concentration is higher but the carrier concentration is lower strongly argues in favour of a carrier mediated exchange [8.42]. Figure 10 shows a schematic of the mechanism where the origin of FM is related to magnetic centres associated with the $V_{Ti}$. As the free electron carrier density of the $Ti_{1-x}Ta_xO_2$ (x~0.05) film is about $7.6 \times 10^{20}$ $cm^{-3}$, the mechanism of FM is most likely facilitated through itinerant electron-mediated RKKY.

In conclusion, the FM seen in $Ti_{1-x}Ta_xO_2$ (x~0.05) thin films prepared at 600 $^o$C is verified by a battery of magnetic measurements and the role of magnetic artefacts was eliminated by a variety of analytical techniques. With close to 100% substitutionality of the Ta, the activation of only 50% of the Ta implied the presence of compensating defects like $V_{Ti}$ and $Ti^{3+}$. Further spectroscopic evidence clearly showed the role of $V_{Ti}$ and not $Ti^{3+}$ leading to a mechanism where magnetic $V_{Ti}$ were helped by RKKY exchange with the free electrons. This is the first demonstration of magnetism arising from cationic vacancies which may pave the way for other novel magnetic phenomena.

## Acknowledgement

The authors would like to thank G. A. Sawatzky, H. Yang, S. B. Ogale, K. Gopinadhan, C. K. Yong, W. Xiao, K. Singal, R. Minqin, T. Osipowicz and F. Watt for their experimental help and fruitful discussions related to this work. This work is supported by NRF-CRP grant "Tailoring Oxide Electronics by Atomic Control" NRF2008NRF-CRP002-024, NUS YIA, NUS cross-faculty grant, MOE MOE AcRF Tier-2 grant (MOE2010-T2-2-121), FRC, BMBF, and DFG. KSY and YD thank the financial support of the National Basic Research Program of China (Program 973, Grant No. 2007CB613302).

**FIGURE CAPTIONS**

**Figure 1**. RBS random and channelling spectra of $Ti_{1-x}Ta_xO_2$ (x~0.05) films grown on $LaAlO_3$ substrate. In the inset, a random RBS spectrum of an identical sample grown on Si substrate.

**Figure 2**. TOF-SIMS spectra from pure and $Ti_{1-x}Ta_xO_2$ (x~0.05) targets and films grown on Si and $LaAlO_3$ substrates along with spectra obtained from corresponding magnetic calibration samples.

**Figure 3**. The XPS analysis of $Ti_{1-x}Ta_xO_2$ (x~0.05) films at the Ti 2p core levels (a) Pure $TiO_2$ film grown at 600 °C and $Ti_{1-x}Ta_xO_2$ (x~0.05) films grown at (b) 600 °C and (c) 750 °C.

**Figure 4**. Magnetic hysteresis loops for pure (black), $Ti_{1-x}Ta_xO_2$ (x~0.05) thin films grown at 600 °C (blue) and 750 °C (red) in oxygen partial pressure of $1\times10^{-5}$ Torr.

**Figure 5**. The absorption coefficient $\mu$ at (a) Ti $L_{2,3}$ edges and (b) O $K$ edge of the pure $TiO_2$ and $Ti_{1-x}Ta_xO_2$ (x~0.05) films grown at 600 and 750 °C where $\mu^+$ and $\mu^-$ are parallel and anti-parallel alignments between the photon helicity and the sample magnetisation direction. The corresponding SXMCD spectra for the (c) Ti $L_{2,3}$ edges and (d) the O $K$ edge.

**Figure 6(a)**. OMCD signals obtained from a $Ti_{1-x}Ta_xO_2$ (x~0.05) film grown at 600 °C showing the dichroism and spin-polarised magnetisation near the optical band gap. The vertical dotted line represents the position of the optical band-gap. (b) Magnetic hysteresis loop from the OMCD measurement showing ferromagnetic behaviour. In the inset, this loop is overlapped with the 40 K SQUID data.

**Figure 7**. The XAS for pure $TiO_2$ and $Ti_{1-x}Ta_xO_2$ (x~0.05) samples grown at (a) 600 °C and (b) 750 °C. The XAS for $Ti_{1-x}Ta_xO_2$ (x~0.05) sample grown at 600 °C shows anomalous enhancement of the spectral weight at $t_{2g}$ states compared to the pure $TiO_2$ grown at same temperature confirming the formation of significant amount of $V_{Ti}$ in $Ti_{1-x}Ta_xO_2$ films. In contrast, the XAS for $Ti_{1-x}Ta_xO_2$ (x~0.05) sample grown at 750 °C shows decrease of the spectral weight at $t_{2g}$ states compared to the pure $TiO_2$ grown at 750 °C showing absence of $V_{Ti}$.

**Figure 8**. Calculated density of states of Ta incorporated anatase $TiO_2$ system: (a) total DOS (b) partial DOS for O $2p$ states. (c) $t_{2g}$ and $e_g$ of Ti $3d$ (d) $t_{2g}$ and $e_g$ of Ta $5d$.

**Figure 9**. Comparison between experimental and calculated OMCD data. The blue line through the OMCD experimental data points is a guide to the eye only.

**Figure 10**. (a) Three-dimensional spin density plot of anatase $TiO_2$ with two $V_{Ti}$. The yellow isosurface represents the spin density of $V_{Ti}$, and dashed green circles show the range of the delocalized magnetic orbitals of $V_{Ti}$ (b) A schematic of the maximum possible ferromagnetic ordering of magnetic centers (gray circle) at the sites of Ti vacancies coupled by itinerant electrons mediated (RKKY) exchange mechanism.

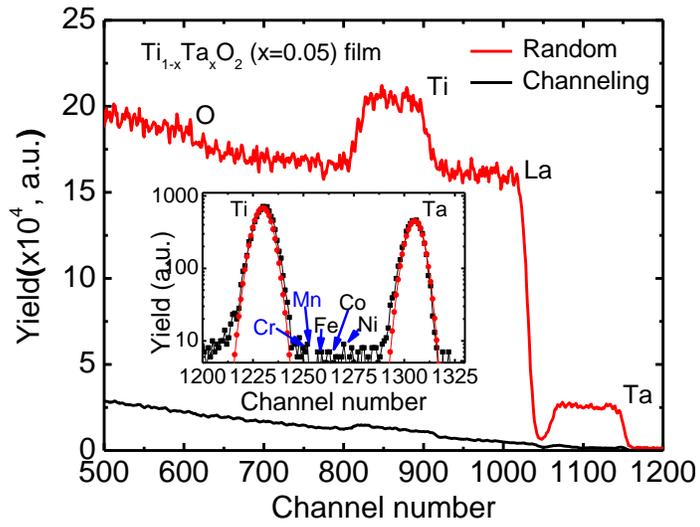

**Figure 1**

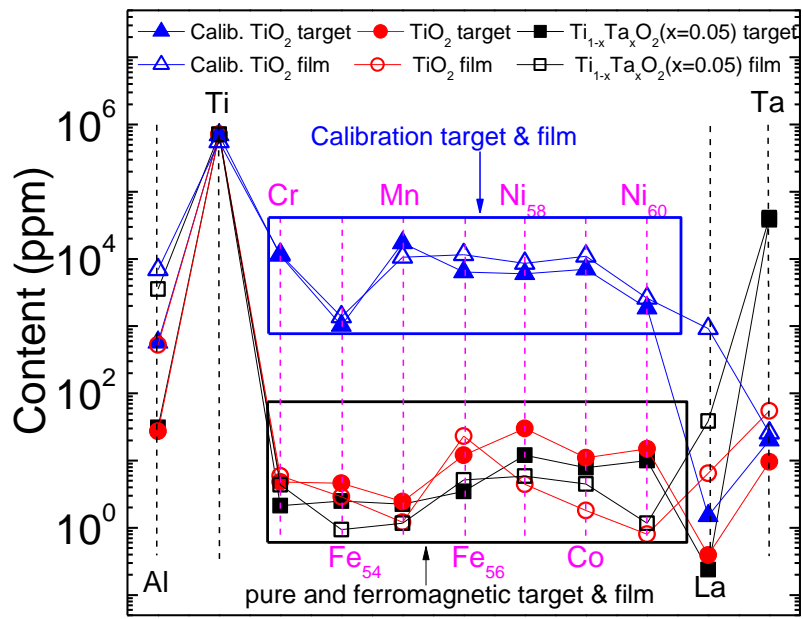

**Figure 2**

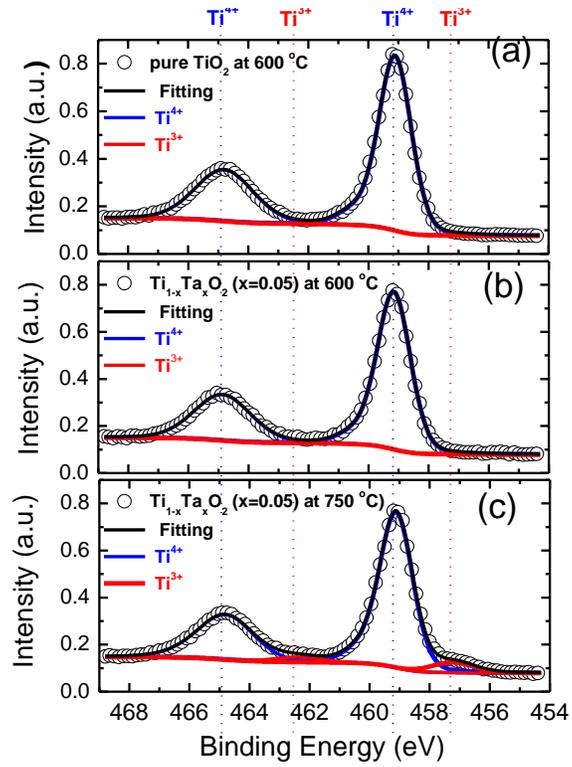

**Figure 3**

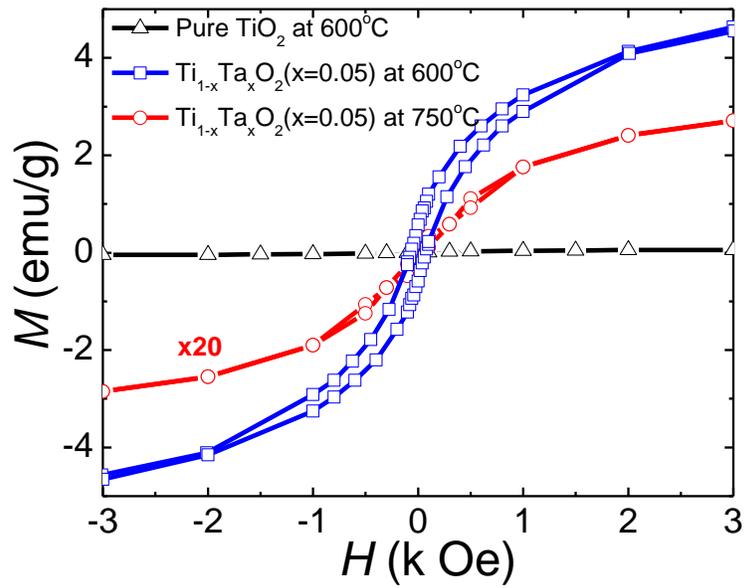

**Figure 4**

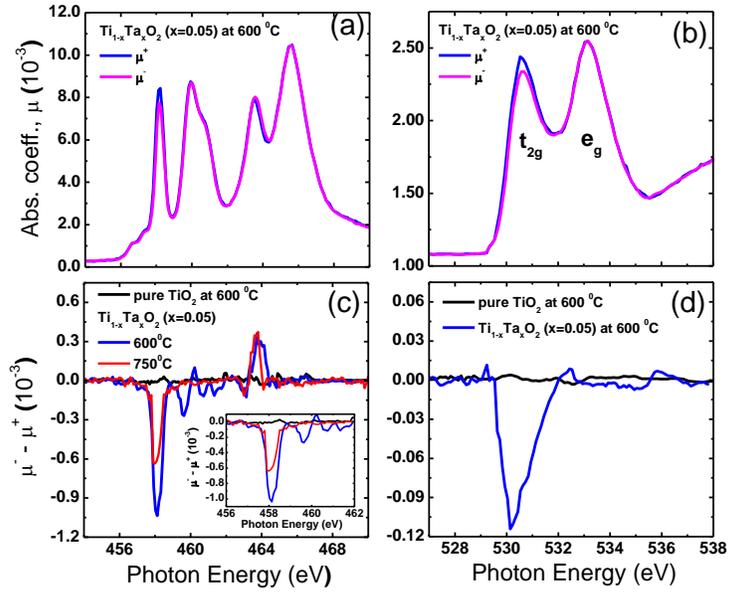

**Figure 5**

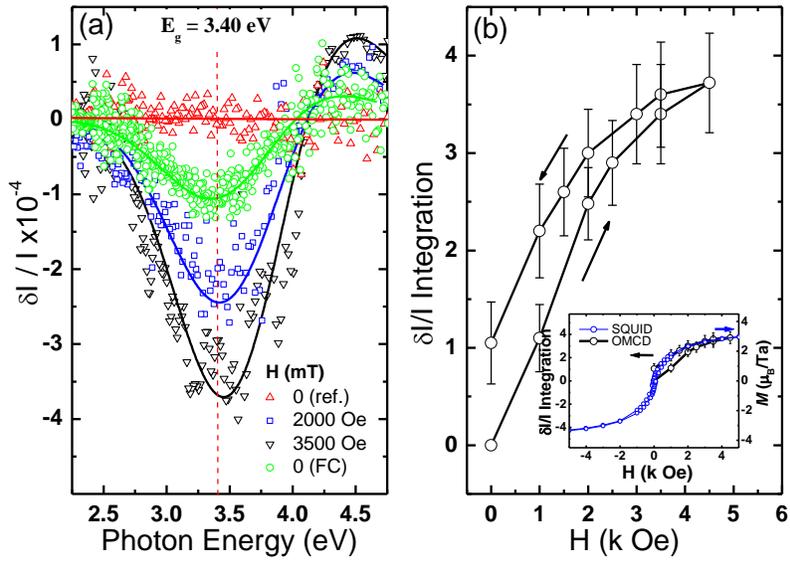

**Figure 6**

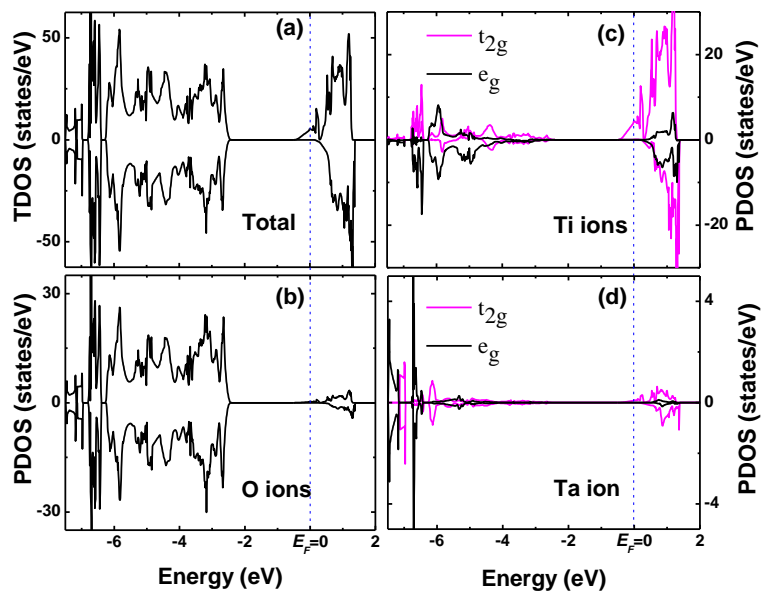

**Figure 7**

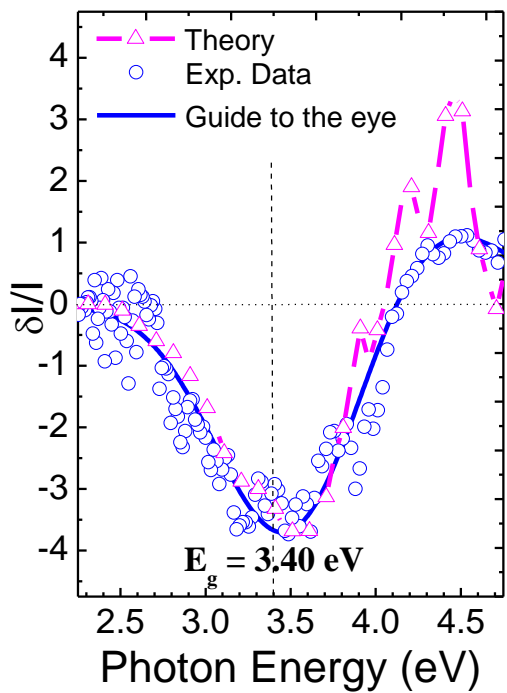

**Figure 8**

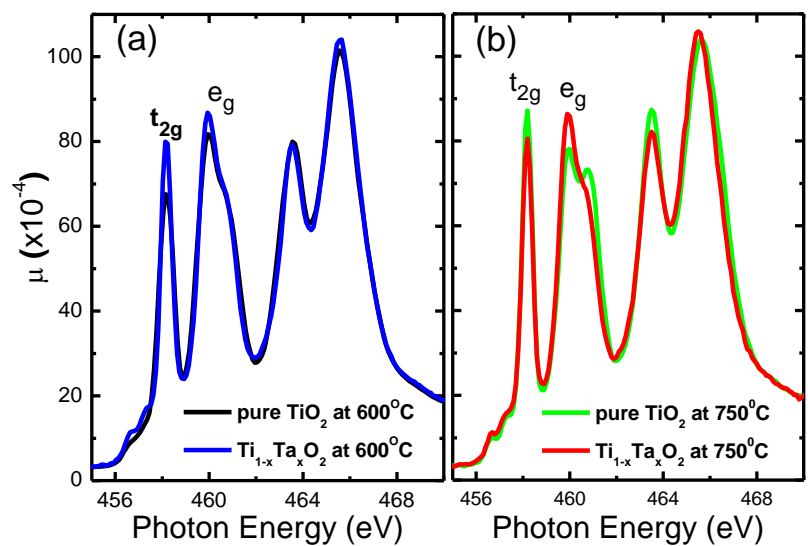

**Figure 9**

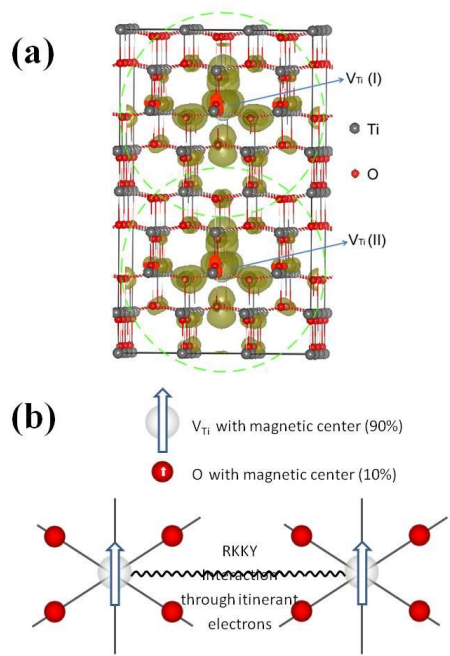

**Figure 10**